\documentclass{ieeeaccess}
\usepackage{cite}
\usepackage{amsmath,amssymb,amsfonts}
\usepackage{algorithmic}
\usepackage{graphicx}
\usepackage{textcomp}
\usepackage{caption}
\definecolor{volrblue}{rgb}{0,0,0.6078}

\def\BibTeX{{\rm B\kern-.05em{\sc i\kern-.025em b}\kern-.08em
    T\kern-.1667em\lower.7ex\hbox{E}\kern-.125emX}}

\begin{document}

\history{This work has been submitted to IEEE for possible publication. Copyright may be transferred without notice, after which this version may no longer be accessible. Date of publication xxxx 00, 0000, date of current version xxxx 00, 0000. \doi{10.1109/ACCESS.2025.DOI}}


\title{Estimating Perceptual Attributes of Haptic Textures Using Visuo-Tactile Data}
\author{\uppercase{Mudassir Ibrahim Awan} and \uppercase{Seokhee Jeon}}
\address{Department of Computer Engineering, Kyung Hee University, South Korea (e-mail: miawan@khu.ac.kr, jeon@khu.ac.kr)}

\tfootnote{This research was supported by the IITP under the Ministry of Science and ICT Korea through the IITP program No. 2022-0-01005 and under the metaverse support program to nurture the best talents ( IITP-2024-RS-2024-00425383).}

\markboth
{Author \headeretal: Preparation of Papers for IEEE Access}
{Author \headeretal: Preparation of Papers for IEEE Access}

\corresp{Corresponding author: Seokhee Jeon (e-mail: jeon@khu.ac.kr).}

\begin{abstract}
Accurate prediction of perceptual attributes of haptic textures is essential for advancing VR and AR applications and enhancing robotic interaction with physical surfaces. This paper presents a deep learning-based multi-modal framework, incorporating visual and tactile data, to predict perceptual texture ratings by leveraging multi-feature inputs. 
To achieve this, a four-dimensional haptic attribute space encompassing rough-smooth, flat-bumpy, sticky-slippery, and hard-soft dimensions is first constructed through psychophysical experiments, where participants evaluate 50 diverse real-world texture samples. 
 A physical signal space is subsequently created by collecting visual and tactile data from these textures.
Finally, a deep learning architecture integrating a CNN-based autoencoder for visual feature learning and a ConvLSTM network for tactile data processing is trained to predict user-assigned attribute ratings. This multi-modal, multi-feature approach maps physical signals to perceptual ratings, enabling accurate predictions for unseen textures.
To evaluate predictive accuracy, we employed leave-one-out cross-validation to rigorously assess the model's reliability and generalizability against several machine learning and deep learning baselines. Experimental results demonstrate that the framework consistently outperforms single-modality approaches, achieving lower MAE and RMSE, highlighting the efficacy of combining visual and tactile modalities.
\end{abstract}


\titlepgskip=-15pt

\maketitle

\section{Introduction}
\label{sec:introduction}
\PARstart{W}{hen} a textured surface is stroked, a range of tactile signals is generated, playing a crucial role in surface texture perception. Humans interpret these signals rapidly within a multi-dimensional haptic attribute space, characterized by descriptors such as roughness, softness, and bumpiness \cite{yoo2016large}. This cognitive process, driven by deformation of the user's skin connected to the user's interactions, enables material recognition and object identification with remarkable accuracy \cite{richardson2022learning, yoo2013consonance}. Computationally modeling this perception is essential for applications in virtual reality (VR), augmented reality (AR), haptic-enabled robotics, and human-computer interaction \cite{takahashi2019deep}.

Primary research on haptic textures has focused on modeling surface interactions and generating realistic haptic feedback. A number of studies have introduced haptic texture rendering modules and libraries that synthesize acceleration signals of specific textures based on physical interactions \cite{abdulali2018data, culbertson2014modeling}. These modules provide a means to generate haptic feedback for virtual environments. However, while the synthesis of texture signals has been well studied, the question of when to use these models effectively remains largely unexplored. Understanding their applicability is crucial for improving haptic rendering fidelity, particularly in scenarios requiring accurate perceptual predictions.

One of the key applications where accurate haptic attribute prediction is essential is model-mediated teleoperation, where a remote system captures interaction data, but real-time transmission of raw haptic signals may be limited due to latency. In such cases, a perceptually aligned haptic texture model can be selected from a texture library \cite{abdulali2018data}, ensuring that the reconstructed feedback on the operator’s side closely matches the intended material properties, thereby enhancing realism in remote interactions \cite{awan2023model}. These systems utilize a rigid tool equipped with vision and tactile sensors, where the vision sensor captures global texture characteristics, such as macro-scale structure and material reflectance, while the tactile sensor records tool-induced vibrations as high-frequency acceleration signals, along with scanning speed and applied force. However, to ensure that the remotely rendered haptic feedback is perceptually consistent with the original material, accurate haptic attribute prediction is needed to establish a reliable mapping between physical signals and human perception. By leveraging predictive models, the system can generate an appropriate haptic representation rather than relying on direct signal transmission that may be affected by latency or interaction variability.

To better conceptualize the idea, we introduce two fundamental representational spaces: the perceptual attribute space and the physical signal space. The perceptual, or haptic, attribute space is constructed through psychophysical experiments, where participants rate textures along bipolar attributes such as rough–smooth and hard–soft, forming a subjective representation of haptic perception \cite{okamoto2012psychophysical, drewing2018systematic}. In contrast, the physical signal space is derived from measured texture characteristics, including high-resolution visual data and tactile signals such as acceleration, applied force, and scanning speed, which collectively encode the objective physical properties of textures \cite{zhang2023visual, strese2016multimodal}. Establishing a reliable mapping between these spaces is essential for computational models to predict how a given texture will be perceived based on its physical attributes.

While humans use both the visual and tactile data to gauge the haptic texture, most previous studies on the prediction of haptic texture perceptual attributes rely on single-modality approaches, using either tactile signals or visual texture analysis. On one hand, while tactile data provide high accuracy in capturing micro-textural properties, it is highly sensitive to interaction parameters, such as force, speed, and sensor noise, which can introduce inconsistencies in estimation. On the other hand, visual data possess macro-scale structural patterns but do not provide compliance-related attributes, such as softness or friction, which are not always visually discernible \cite{richardson2022learning, awan2023model}. Visual data also do not have micro-scale information that greatly influences texture perception, due to the spatial resolution of the visual sensors. Some studies have attempted to integrate multi-modal learning, but they predominantly focus on texture classification rather than continuous haptic attribute prediction, limiting their ability to model perceptual variations accurately \cite{zhang2023visual, chen2023research, awan2023model}. Since vision and tactile data encode complementary information, a robust multi-modal framework that effectively fuses both modalities is necessary to improve haptic attribute prediction, enhance generalizability, and ensure perceptual alignment with human ratings.

Nonetheless, while various computational techniques have been explored to map perceptual attributes from physical signals, challenges remain in achieving robust and generalizable haptic prediction. Early attempts employed parametric models but often struggle to generalize across different textures and interaction conditions \cite{richardson2022learning}. More recent efforts have leveraged deep learning-based models to learn complex mappings between input signals and haptic attributes \cite{awan2023predicting, hassan2023establishing}. These methods have demonstrated notable success, particularly in capturing intricate texture representations and improving prediction accuracy. However, many existing deep learning models are still trained on single-modality data, which can limit their ability to generalize across diverse textures. Multi-modal deep learning approaches that integrate visual and tactile data have shown promising results, with different fusion strategies being explored to enhance their effectiveness. One promising direction involves cross-modal feature fusion, where representations extracted from different sensory modalities are effectively combined to improve prediction accuracy \cite{zhang2023visual, awan2023model}. Additionally, leveraging features extracted from pre-trained models and integrating them with classical handcrafted descriptors provides a robust way to capture both high-level abstract features and fine-grained physical properties, further enhancing the reliability of haptic attribute prediction \cite{hassan2023establishing}.

Motivated by these challenges, this work introduces a deep learning framework that integrates visual and tactile data for predicting perceptual haptic attributes. The framework constructs a physical signal space by capturing high-resolution images and tactile data, including acceleration, applied force, and scanning speed from 50 real-world textures, while simultaneously establishing a four-dimensional perceptual space through psychophysical experiments, where participants rate textures along the bipolar attributes of rough-smooth, flat-bumpy, sticky-slippery, and hard-soft. To facilitate the analysis of these perceptual ratings, the four-dimensional space is visualized in a two-dimensional representation, providing insights into the structure of human haptic perception. 
To establish a mapping between these spaces, the framework employs a two-stream architecture, where the visual stream extracts texture features using a CNN-based autoencoder, incorporating pre-trained ResNet-50 \cite{he2016deep} features alongside Gray-Level Co-occurrence Matrix (GLCM) descriptors to enhance structural representation. The tactile stream, implemented as a Convolutional LSTM (ConvLSTM) network \cite{shi2015convolutional}, processes high-frequency vibration signals using Mel-Frequency Cepstral Coefficients (MFCCs), complemented by interaction parameters such as scanning speed and applied force to improve robustness. By integrating these complementary modalities, the framework strengthens feature representation and enhances perceptual alignment, leading to more accurate haptic attribute predictions.

Beyond applications in teleoperation, haptic attribute prediction can also serve as a scalable alternative to human perceptual evaluation. Directly assessing texture properties through psychophysical experiments is often impractical due to time, cost, and logistical constraints, while in certain cases, such as analyzing hazardous surfaces or conducting large-scale material perception studies, direct human interaction is infeasible \cite{okamoto2012psychophysical, drewing2018systematic}. By leveraging multimodal sensory data, a predictive model can enable efficient, reproducible, and scalable estimation of perceptual haptic attributes, reducing the dependency on resource-intensive human studies while maintaining perceptual alignment.
Another key application is perception-based data compression and transmission. Instead of storing and transmitting raw physical data, perceptual attributes can be estimated from newly captured signals, encoded with compression, and efficiently stored or transmitted. This approach can significantly reduce data storage requirements and transmission bandwidth. Eventually, rendering algorithms and haptic devices can convert perceptual attribute values into appropriate commands or physical signals tailored to the user’s interaction.

\begin{figure*} [t]
        \centering
        \includegraphics[width=1.0\textwidth] {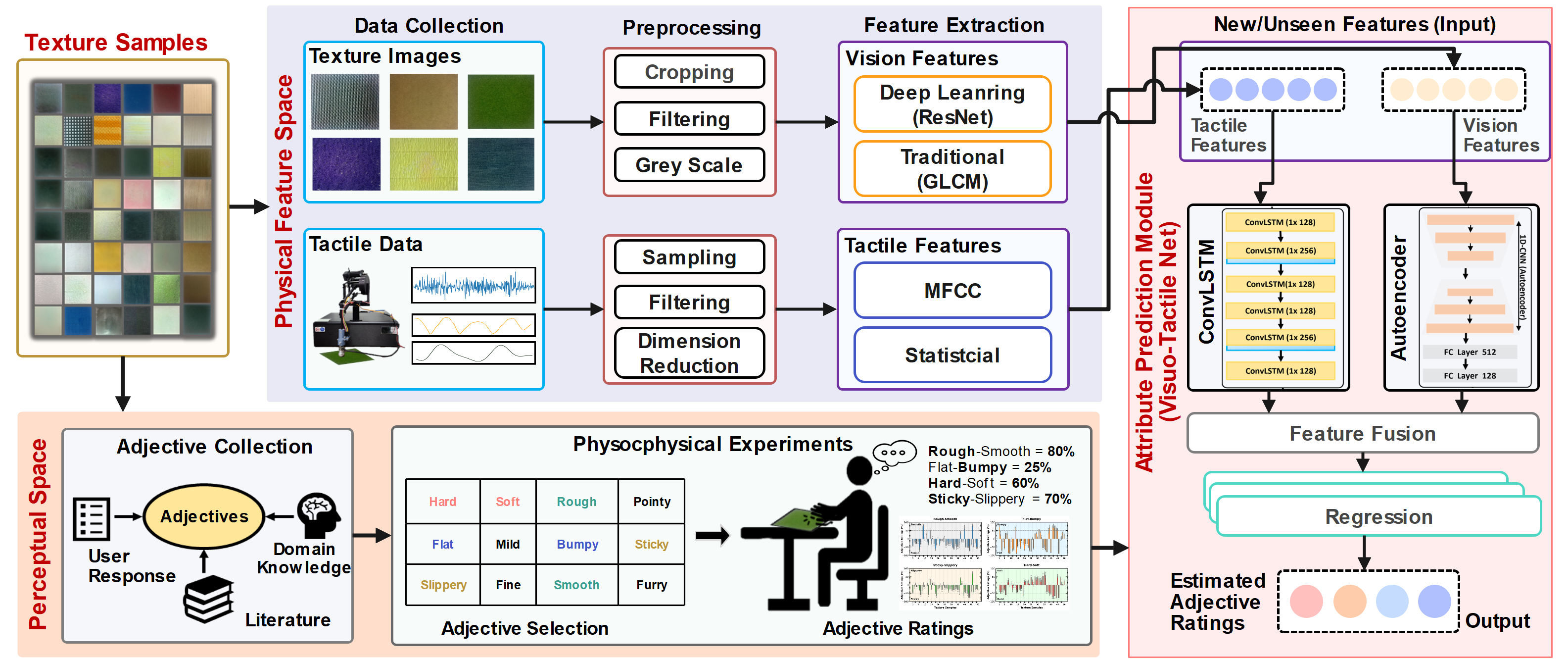}
        \caption{Overall Framework.}
        \label{fig:frameowrk}
\end{figure*}

The paper is organized as follows. Figure \ref{fig:frameowrk} illustrates the overall framework. Section \ref{sec:background} provides a review of related work. The proposed method, including the architecture of the attribute prediction model and its input-output schema, is introduced in Section \ref{sec:propsoed_method}. Section \ref{sec:haptic_perceptual_space} describes the construction of the haptic perceptual space.  The collection and preprocessing of visual-tactile data, which form the basis of the physical feature space, are outlined in Section \ref{sec:Visio_tactile_dataset_collection}.  Evaluation procedures and results are presented in Section \ref{sec:evaluation_experiments}, followed by a discussion of the framework in Section \ref{sec:discussion}. Finally, Section \ref{sec:conclusion} concludes the study.

\section{Background}
\label{sec:background}
Below, we discuss related work on haptic texture attributes and their organization within perceptual spaces, the use of tactile and visual data, as well as deep learning approaches for texture analysis.

\subsection{Haptic Texture Attributes and Perceptual Spaces}
\label{subsec:rw_texture_attributes}

Haptic texture attributes are perceptual qualities humans associate with surfaces, such as roughness, and slipperiness. These attributes can be perceived through bare-finger or tool-based interaction and form the basis for constructing haptic perceptual spaces, multidimensional representations that characterize textures by human perception \cite{yoshioka2007texture}.

One of the pioneering studies on understanding haptic texture perception was conducted by Yoshida et al \cite{yoshida1968dimensions}, focusing on bare-finger interactions. Their research identified four key perceptual dimensions of texture: hard-soft, heavy-light, cold-warm, and rough-smooth. 
Subsequent work refined these findings, confirming rough–smooth and hard–soft as dominant bipolar dimensions~\cite{hollins2005factors}. Further expansion introduced macro- and micro-roughness as distinct perceptual axes and highlighted friction as a critical attribute~\cite{gescheider2005perception}. 
Collectively, these studies contributed to the widely recognized five perceptual dimensions: micro-roughness, macro-roughness, friction, stiffness, and warmth.
In contrast, tool-mediated methods, such as those employed by \cite{lamotte2000softness}, demonstrated that tapping with a rigid probe enhances the perception of hardness and softness. 
Other studies demonstrated that tool-based interactions reliably assess the rough–smooth dimension by reducing variability in skin contact~\cite{yoshioka2007texture, culbertson2017ungrounded}. 
Despite their effectiveness, tool-mediated approaches may fail to capture finer details like friction and micro-roughness. In these cases, bare-finger interactions provide richer and more nuanced feedback, which is essential for accurately capturing subtle surface properties \cite{hassan2016evaluating}.

Haptic attributes derived from user interactions are commonly used to construct perceptual spaces that characterize the multi-dimensional nature of texture perception. These spaces are typically generated using techniques like Multi-Dimensional Scaling (MDS) \cite{yoshioka2007texture} or Principal Component Analysis (PCA) \cite{baumgartner2013visual,chu2015robotic}, which reduce dimensionality for easier interpretation. Perceptual spaces play a crucial role in texture analysis \cite{wu2015experimental,culbertson2014modeling} and the development of virtual textures \cite{hassan2019authoring}.
While dimensionality reduction simplifies data, it can overlook important perceptual details and is unsuitable for estimating actual human-assigned ratings. In our recent work \cite{hassan2023establishing}, we introduced a four-dimensional Haptic Attribute Space consisting of two 2D subspaces, preserving raw user ratings without reducing dimensions. This approach offers a more accurate and detailed representation of perceptual attributes, directly reflecting the degree of user-assigned ratings for each texture dimension.
Despite progress, further research is needed to develop intuitive representations that incorporate actual user ratings, enhancing the understanding of haptic perception.

\subsection{Tactile and Vision Data for Texture Analysis}
\label{subsec:rw_visual_tactile}

Texture analysis through tactile feedback involves capturing the unique vibrations generated when interacting with surfaces. This feedback reflects both macro features (e.g., bumpiness) and micro features (e.g., fine roughness) \cite{lamotte2000softness, hassan2016evaluating}. However, the relationship between texture properties, user motion, and the resulting vibrations is inherently complex and nonlinear, posing significant challenges for accurate modeling and distinguishing \cite{strese2015surface}.  Early studies recorded tactile data under fixed interaction parameters or segmented it into stationary signals, limiting generalizability \cite{awan2023model}. Recent approaches have shifted towards directly utilizing data collected through free-hand motion,  preserving natural interactions and a wider range of vibratory responses \cite{lu2022preference} without information loss. However, even with parametric and deep learning models, distinguishing similar textures remains difficult due to overlapping vibratory signals \cite{richardson2022learning,lu2022preference,awan2023predicting}.

In contrast,Vision-based techniques offer a simpler alternative to tactile sensors, requiring less specialized hardware. Heravi et al. \cite{heravi2024development} used GelSight images to classify textures effectively, though their focus was on texture rendering rather than precise attribute prediction. \cite{yang2024binding} aligned GelSight embeddings with images and audio to enhance texture classification. 
While \cite{hassan2023establishing} employed a feature-based approach using texture images to estimate haptic attributes, showing strong results but struggling in predicting attributes like softness and fine roughness. This is likely because image-based methods primarily capture macro features (e.g., surface patterns) but often miss sub-surface features (e.g., material compliance / softness), limiting their accuracy in similar applications.

These challenges are well-recognized in the haptics community. To address them, studies have explored integrating visual and tactile features for more robust texture analysis \cite{baumgartner2013visual}. Fusing visual data, which captures macro features, with tactile data, which reflects micro details, enables a comprehensive representation of textures. This multi-modal approach improves the prediction of perceptual attributes, surpassing simple texture classification \cite{strese2016multimodal, lin2024object}. By leveraging shared features from both modalities, models achieve better generalization and accuracy, even for unseen textures. However, most efforts focus on classification, with limited research addressing regression for predicting perceptual haptic attributes \cite{li2024classification, strese2016multimodal}.

\begin{figure*} [t]
        \centering
       \includegraphics[width=0.98\textwidth] {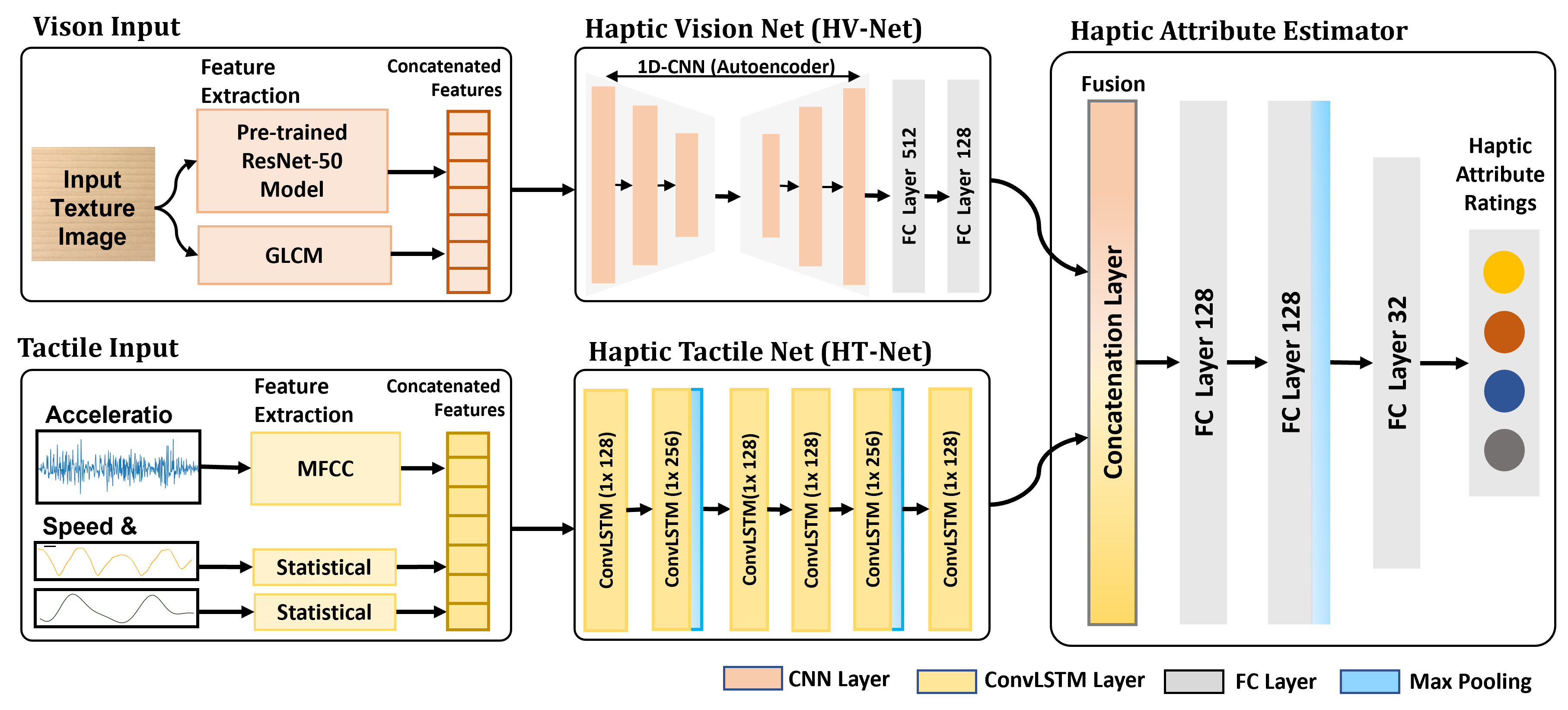}
        \caption{The proposed visuo-tactile network. It consists of two streams: one for visual data employing an autoencoder and another for tactile data utilizing a 1D CNN.}
        \label{fig:model}
\end{figure*}

\subsection{ Deep Learning Approaches for Texture Perception}
\label{subsec: Deep Learning Approaches for Texture Perception}

Recent studies have increasingly adopted deep learning (DL) approaches for modeling texture perception from both visual and tactile data \cite{lin2024object,awan2023predicting}. Convolutional Neural Networks (CNNs) are commonly used for extracting spatial features from visual textures, effectively capturing structural and geometric patterns in images, and have shown significant performance in texture recognition tasks \cite{hassan2023establishing}. Tactile signals, in contrast, are inherently spatio-temporal, as they contain both local surface-related information and dynamic variations over time.
In earlier works, researchers applied CNN-based models to time-series tactile data, focusing on extracting local features from vibration signals \cite{strese2016multimodal}. With the introduction of Recurrent Neural Networks (RNNs), particularly Long Short-Term Memory (LSTM) networks that are specialized in modeling temporal dependencies, researchers began to explore CNN-LSTM hybrid architectures. These models typically combine CNN and LSTM in a stacked or two-stream configuration, where CNNs are used to extract spatial features from input segments and LSTMs capture temporal dynamics across those segments \cite{awan2023predicting}. While this setup enables joint spatio-temporal learning, it often introduces challenges including complex optimization, sensitivity to hyperparameters, and reduced spatial coherence when features are temporally sequenced. These limitations are not exclusive to haptic data and have been widely observed in other spatio-temporal learning tasks \cite{awan2023predicting,byeon2014texture,hassan2024quantifying}.

To address the limitations of existing architectures for processing time-series data, recent works have explored several alternatives, with Transformer frameworks \cite{vaswani2017attention} and Convolutional LSTM (ConvLSTM) networks \cite{shi2015convolutional} being among the most prominent. Transformers perform well in sequential tasks but are constrained by large data needs and can limit their practicality in texture-based haptic tasks \cite{wen2022transformers}.ConvLSTM, however, offers a more structured approach for modeling both spatial and temporal dependencies and has achieved strong results in various time-series applications, including water irrigation forecasting \cite{bounoua2024deep} and surface deformation prediction \cite{yao2023convlstm}. Its ability to jointly preserve spatial structure while learning temporal dynamics makes it particularly suitable for dense, sensor-based sequences such as tactile signals.

Despite its success in related domains, ConvLSTM remains unexplored for haptic texture analysis. We hypothesize that it is better suited for learning the underlying spatial and temporal structure of tactile signals compared to LSTMs or CNN-LSTM hybrids.For the visual modality in our multimodal framework, we consider CNNs an effective choice for extracting spatial features, including local texture patterns and geometric structures, given their proven ability to capture these characteristics in texture images.

\section{Attribute Prediction Module}
\label{sec:propsoed_method}
\label{subsec:model_architecture}

The primary objective of this work is to predict haptic affective attributes from multimodal physical signals using a structured computational framework. As shown in Figure~\ref{fig:frameowrk}, the process begins with the preparation of texture samples obtained from real-world surfaces. Next, we construct two distinct data spaces: 1) Physical Feature Space (PFS), which includes multimodal physical signals captured from the texture samples, and 2) Haptic Perceptual Space (HPS), which contains user-assigned perceptual attribute ratings gathered through psychophysical experiments.
The core of this framework is the Attribute Prediction Module (APM), which bridges the gap between physical signals and human perceptual ratings. The APM design is detailed in the remainder of this section, while the construction of the HPS and PFS is described in Sections~\ref{sec:haptic_perceptual_space} and~\ref{sec:Visio_tactile_dataset_collection}, respectively.

To facilitate the mapping between physical features and perceptual attributes, the APM employs a dual-stream architecture which we also termed as Visuo-tactile Net (Figure~\ref{fig:model}) consisting of two parallel branches: the Haptic Vision Network (HV-Net) and the Haptic Tactile Network (HT-Net). Both networks operate on pre-extracted features rather than raw data to improve robustness and mitigate overfitting (see Sections~\ref{sec:Visio_tactile_dataset_collection} and~\ref{sec:haptic_perceptual_space}). HV-Net encodes visual information from texture images, while HT-Net models temporal patterns in tactile signals. Finally, the dual-stream architecture fuses visual and tactile features from HV-Net and HT-Net to create a robust joint representation of physical texture, which is then used to predict haptic attributes. The following subsections describe the design of each stream and the associated training methodology.

\subsection{Haptic Vision Network (HV-Net)}  

The HV-Net generates compact and discriminative representations from visual texture inputs by integrating deep and statistical features. The input to HV-Net combines high-level descriptors extracted using a pretrained ResNet-50 model~\cite{he2016deep} with handcrafted texture descriptors derived from the Gray-Level Co-occurrence Matrix (GLCM). A detailed description of these visual feature extractions is provided in Sec. ~\ref{subsec:Image_dataset_collection}.

To process this high-dimensional input while mitigating overfitting and preserving relevant structure, HV-Net employs a convolutional autoencoder (CNN-AE) built on 1D convolutional layers.  The use of 1D-CNNs, instead of 2D-CNNs, is motivated by the nature of the input, which is a flattened feature vector without spatial dimensions. This choice significantly reduces the number of trainable parameters, improves generalization, and allows the model to capture local patterns efficiently. The encoder consists of sequential 1D-CNN layers with filter sizes of 256, 256, 128, 64, and 32, and kernel sizes of $(1 \times 4), (1 \times 4), (1 \times 3), (1 \times 3),$ and $(1 \times 3)$, respectively. Each convolutional layer is followed by a max pooling operation with a pooling size of $(1 \times 2)$ to reduce temporal resolution and improve robustness. The decoder mirrors this structure in reverse order, applying 1D-CNN layers with filter sizes of 32, 64, 128, 256, and 256 to reconstruct the original input feature vector. This self-supervised reconstruction allows the network to learn stable and discriminative visual features by suppressing irrelevant variations while retaining meaningful texture structure.

The output of the decoder is passed through two fully connected layers with 512 and 128 units, respectively, each followed by a ReLU activation. This projection compresses the learned representation into a compact 128-dimensional visual feature vector, which is later combined with the tactile features from HT-Net during the multimodal fusion stage.

\begin{figure*} [t]

\centering
        \includegraphics[width=0.98\textwidth] {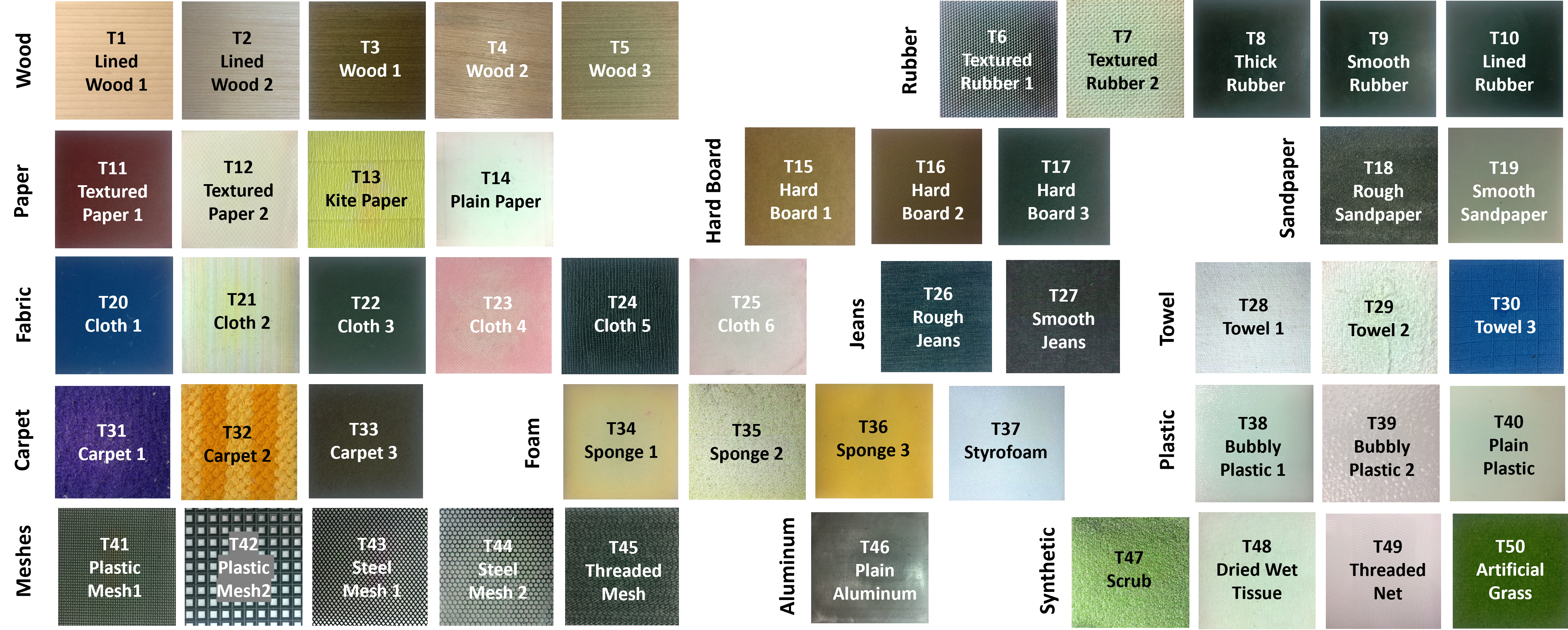}
        \caption{The real texture samples used in this study from diverse categories}
        \label{fig:texture_dataset}
\end{figure*}

\subsection{Haptic Tactile Network (HT-Net)}  

During surface exploration, stroking motions involve user-controlled interaction parameters such as scanning speed $(v)$ and applied force $(f)$, which determine how the surface is explored. These interactions produce vibrations that reflect surface properties including microstructure, roughness, and friction. The resulting dynamic responses are captured through acceleration signals $(a)$, which contain high-frequency components induced by contact with the surface. Acceleration signals are often affected by sensor noise and variability in hand motion, particularly under unconstrained conditions. To improve robustness, HT-Net operates on extracted features rather than raw signals. The continuous tactile recordings are first segmented into overlapping temporal windows, with each segment treated as a single input instance. For the acceleration signal, Mel-Frequency Cepstral Coefficients (MFCCs) are computed to capture spectral content in a compact, noise-resilient form. In contrast, scanning speed and applied force are low-frequency signals that exhibit limited variation within short intervals. Therefore, statistical descriptors are computed from the speed and force signals to summarize their temporal behavior across each segment. The complete feature extraction process is described in Sec. \ref{subsec:tactile_dataset_collection}.

For each temporal segment, a feature vector is defined as
\[
X_t = (\text{MFCC}_a, \text{statistical}(v), \text{statistical}(f)),
\]
where $\text{MFCC}_a$ denotes cepstral features extracted from $a$, and $\text{statistical}(v)$ and $\text{statistical}(f)$ represent statistical descriptors computed separately from $v$ and $f$. Each segment-level vector $X_t$ forms one input in the sequence provided to the ConvLSTM. The components are concatenated into a single input vector per segment.

HT-Net uses a ConvLSTM-based architecture to effectively capture both local spatial structure and long-range temporal dependencies present in sequential tactile signals. ConvLSTM combines convolutional operations with recurrent memory, making it particularly suitable for spatio-temporal modeling of tactile data~\cite{shi2015convolutional}. The network comprises six stacked 1D-ConvLSTM layers with filter sizes of 128, 256, 128, 128, 256, and 128, respectively. To reduce temporal resolution and enhance representational efficiency, temporal max pooling operations with a window size of $1 \times 2$ are applied after the 2nd, 4th, and 5th ConvLSTM layers. This layer-wise architecture enables HT-Net to extract hierarchical tactile representations while progressively compressing the temporal dimension. The final hidden state output forms a compact 128-dimensional tactile representation, which is later fused with the visual stream for joint haptic attribute prediction. Each ConvLSTM layer follows the original formulation introduced in~\cite{shi2015convolutional} and is implemented using TensorFlow Keras~\cite{tensorflowkeras}.

\subsection{Output and Training Method}
\label{subsec:training_method}

The final visual and tactile representations from HV-Net and HT-Net, each 128-dimensional, are concatenated to form a 256-dimensional multimodal feature vector. This vector passes through two fully connected (FC) layers with 128 units, followed by a max pooling layer with a window size of $1 \times 2$. The pooled output is processed by an FC layer with 32 units and a final FC layer with 4 output neurons, which predict the haptic attribute scores.

The configuration of the overall architecture, including the number of layers, filter sizes, and fully connected dimensions, was determined through extensive empirical experiments. The network is trained end-to-end using the TensorFlow-Keras framework with the Adam optimizer and RMSE loss. ReLU activation is used in all intermediate layers, while the final output layer employs linear activation to support continuous regression. Training is performed for up to 200 epochs, with early stopping based on validation performance and a patience of 10 epochs.

\section{Haptic Perceptual Space (HPS)}
\label{sec:haptic_perceptual_space}

This section briefly describes the process of creating the proposed Haptic Perceptual Space (HPS) using human participants. The first step involves conducting a psychophysical experiment to identify texture attributes that characterize perceptual properties. In the second part of the experiment, participants rated the texture attributes identified in the prior phase. It is noted that the experimental setup and dataset were adopted from our previous study \cite{hassan2023establishing}. Followings provide the details of the whole process briefly:

\subsection{Texture Dataset}
\label{subsec:texture_dataset}
This study utilizes 50 real texture samples from diverse categories to construct both the Haptic Perceptual Space (HPS) and the Physical Signal Space (PSS). The textures were carefully selected to represent a broad spectrum of materials and surface properties.
To ensure comprehensive coverage, the dataset was categorized into 16 distinct classes. Each class contains textures exhibiting various characteristics, including differences in roughness, softness, slipperiness, and other tactile properties.
The texture categories include wood, rubber, paper, hardboard, sandpaper, fabric, jeans, towels, carpet, foam, plastic, meshes, aluminum, and synthetic materials. A detailed overview of the 50 texture samples is presented in Figure \ref{fig:texture_dataset}.

Furthermore, all the texture samples were cut to 100x100 mm for standardization. They were then affixed to pre-prepared hard acrylic plates of the same dimensions. The acrylic plates, measuring 100x100x5 mm, ensured uniform surface elevation across all samples. Liquid surface glue was used to attach the textures securely. 
This mounting process was implemented to ensure uniform surface elevation across all samples and to prevent any unevenness or curling of the texture during the experiments\cite{awan2023predicting,hassan2023establishing}.

\subsection{Experiment 1: Attribute Selection}
\label{subsec:attribute_selection}

The initial phase of this psychophysical study focused on identifying key adjectives that describe human perception of surface textures upon interaction.
For this study, we gathered 60 haptic texture-related attributes/adjectives that can represent the dataset. A total of three sources were used to gather these adjectives: literature \cite{ yoo2016large, yoo2013consonance}, domain knowledge, and user experiments. The full list of adjectives used during the experiment is shown in Table \ref{tab:attributes_list}.


\begin{table}
\centering
\caption{Attributes presented to participants for the attribute selection experiment. The four selected attribute pairs, highlighted in bold dark blue, were subsequently used for the rating experiment.}
\label{tab:attributes_list} 
\begin{tabular}{|c|c|c|c|c|c|}
\hline
Refined                              & Jarred                      & Bald                           & Mushy   & {\color{volrblue} \textbf{Flat}} & Vague           \\ \hline
{Furry} & {Grating} & {Silky} & Warm    & Thick        & {\color{volrblue} \textbf{Smooth}}           \\ \hline
{\color{volrblue} \textbf{Hard}} & {Bouncy} & {Pleasant} & Glassy    & Pointy        & Blur           \\ \hline
{\color{volrblue} \textbf{Sticky}}                      & Sharp                       & Dense                          & Angular & Hatched       & Even            \\ \hline
Jagged                               & Spongy                      & {\color{volrblue} \textbf{Bumpy}}                 & Cold    & Slow          & Dark            \\ \hline
Grainy                               & Patterned                   & {\color{volrblue} \textbf{Slippery}}              & Light   & Slick         & Granular        \\ \hline
Distinct                             & Irritating                  & Wooden                         & Mild    & Bright        & {\color{volrblue} \textbf{Rough}}  \\ \hline
Prickly                              & Metallic                    & Bubbly                         & Deep    & Fast          & Heavy           \\ \hline
Solid                                & Fine                        & Blur                           & Shallow & Rigid         & {\color{volrblue} \textbf{Soft}}   \\ \hline
Glassy                               & Thin                        & Hatched                        & Sparse  & Blunt         & Fizzy           \\ \hline
\end{tabular}
\end{table}

\begin{figure*} [t]
        \centering
        \includegraphics[width=0.98\textwidth] {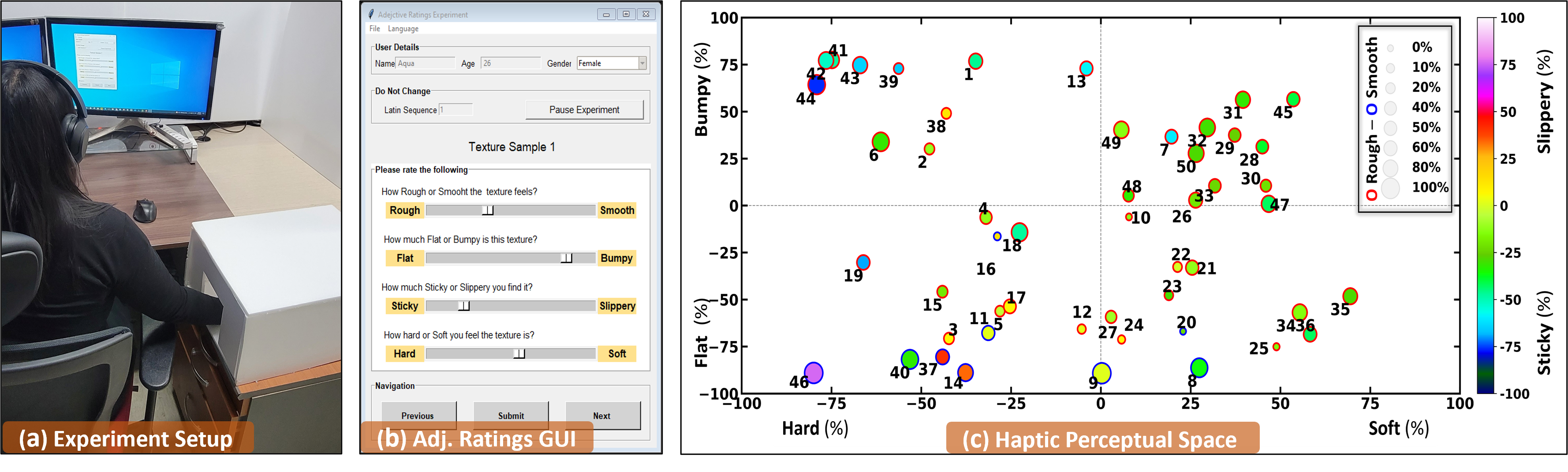}
        \caption{(a) The perceptual experiment setup. (b) The GUI for adjective ratings experiment. (c) Four-dimensional haptic perceptual space visualized as a 2D bubble plot. The plot shows ratings from four adjective pairs for each texture: hard-soft (x-axis), flat-bumpy (y-axis), rough-smooth (bubble size), and sticky-slippery (color gradient). Ratings range from -100 to 100, with -100 representing one extreme (e.g., hard) and 100 the opposite (e.g., soft).}
        \label{fig:perceptual_space}
        
\end{figure*}

\vspace{1em}
\noindent \textbf{Participants:}
A total of 26 participants (19 male and 7 female) took part in this study, with ages ranging from 25 to 34 and an average age of 28. 
All participants were right-handed, used their dominant hand during the experiment, and reported no disabilities that might impact their performance or require special accommodations. 

\vspace{1em}
\noindent \textbf{Experiment Setup:}
The experimental setup is illustrated in Figure \ref{fig:perceptual_space}. Participants were seated at a table, wearing headphones that emitted white noise to minimize environmental distractions. A cardboard box with two openings was placed on the table. One opening featured a small aperture through which participants could insert their hand to explore the textures, effectively blocking visual input during the task. The second opening allowed the experimenter to interchange the textures without revealing them to the participant.
Participants received instructions in both written and verbal form, detailing the task of selecting adjectives to describe the perceived textures.

\vspace{1em}
\noindent \textbf{Stimuli and Procedure:}
The primary objective of this experiment is to identify adjectives that characterize the perception of texture. Each participant was presented with 50 texture samples (see Fig. \ref{fig:texture_dataset}), one at a time, and allowed to explore them freely through touch without time constraints, using any preferred exploratory movements.
Participants evaluated each texture individually and selected adjectives from the provided list (see Table \ref{tab:attributes_list}) that they considered relevant. Their decisions were recorded in binary form: "1" for relevant adjectives and "0" for irrelevant ones.

\vspace{1em}
\noindent \textbf{Results:}
The analysis revealed key attributes that consistently described the texture surfaces. The scores assigned to each adjective across all textures and participants were summed and normalized to generate a relevance score. Adjectives with relevance scores of 50$\%$ or higher were retained for further analysis, yielding a selection of 11 adjectives. From this set, antonymous pairs were identified to represent opposing ends of perceptual dimensions. Adjectives without corresponding antonyms were excluded. The final set consisted of four antonymous pairs: rough–smooth, flat–bumpy, sticky–slippery, and hard–soft.These pairs were used in the next phase of the experiment.

\subsection{Experiment 2: Adjective Ratings}
\label{subsec:adjective_ratings}

\vspace{1em}
\noindent \textbf{Experiment Setup:}
In the second phase, participants rated each texture using antonymous attribute pairs identified in the first experiment. The ratings were recorded through a custom-designed user interface displayed on a PC (see Fig. \ref{fig:perceptual_space}). This interface featured four sliders, each representing one of the antonymous pairs.
The physical length of each slider was 127 mm, following the standardized method  \cite{schiffman1981introduction}. This design ensured sufficient resolution for participants to express subtle perceptual differences, enhancing the precision of data collection in perceptual scaling experiments while maintaining ease of use \cite{schiffman1981introduction, hwang2010perceptual}.
Participants explored each texture with their dominant hand, taking as much time as needed to reach a confident assessment. 
Slider values ranged from 0 to 100, with each slider representing a scale between two opposing attributes displayed at either end, while the numerical values remained hidden from participants.

\vspace{1em}
\noindent \textbf{Results:}
The responses from all participants were aggregated to derive the final perceptual ratings for each texture. For enhanced analysis and visualization, these ratings were averaged and mapped onto a scale ranging from -100 to 100, with 0 representing the midpoint. On this scale, -100 and 100 correspond to the extremes of each attribute (e.g., rough to smooth), with polarity indicating the shift toward opposing haptic properties.

The final outcome of this study is the average rating for each attribute corresponding to each texture, which will be used to map physical signal space to perceptual space, as described in Section  \ref{sec:propsoed_method}. 
To visualize this four-dimensional dataset, we developed a haptic perceptual space (HPS) using a bubble chart with a color gradient. The HPS encodes four dimensions: hard-soft (x-axis), flat-bumpy (y-axis), rough-smooth (bubble size), and sticky-slippery (color gradient).
The HPS plot is illustrated in Fig. \ref{fig:perceptual_space}. To the best of our knowledge, this HPS is the first visualization to consolidate multi-dimensional haptic attributes into a unified 2D framework to display absolute ratings. Unlike previous studies that required multiple graphs to represent each dimension separately \cite{hassan2023establishing}, this approach integrates all sensory dimensions within a single plot, streamlining  the interpretation of texture properties and enabling efficient analysis of large datasets.

\section{Physical Feature Space (PFS)}
\label{sec:Visio_tactile_dataset_collection}

This section defines the Physical Feature Space (PFS), a dataset composed of synchronized tactile signals and visual observations. The first part describes the tactile data, including the hardware configuration, signal acquisition, preprocessing, and feature computation. The second part outlines the visual data, covering image capture and the extraction of both deep and classical texture descriptors.

\subsection{Tactile Dataset }
\label{subsec:tactile_dataset_collection}

\begin{figure} [t]
\centering
        \includegraphics[width=1.0\columnwidth] {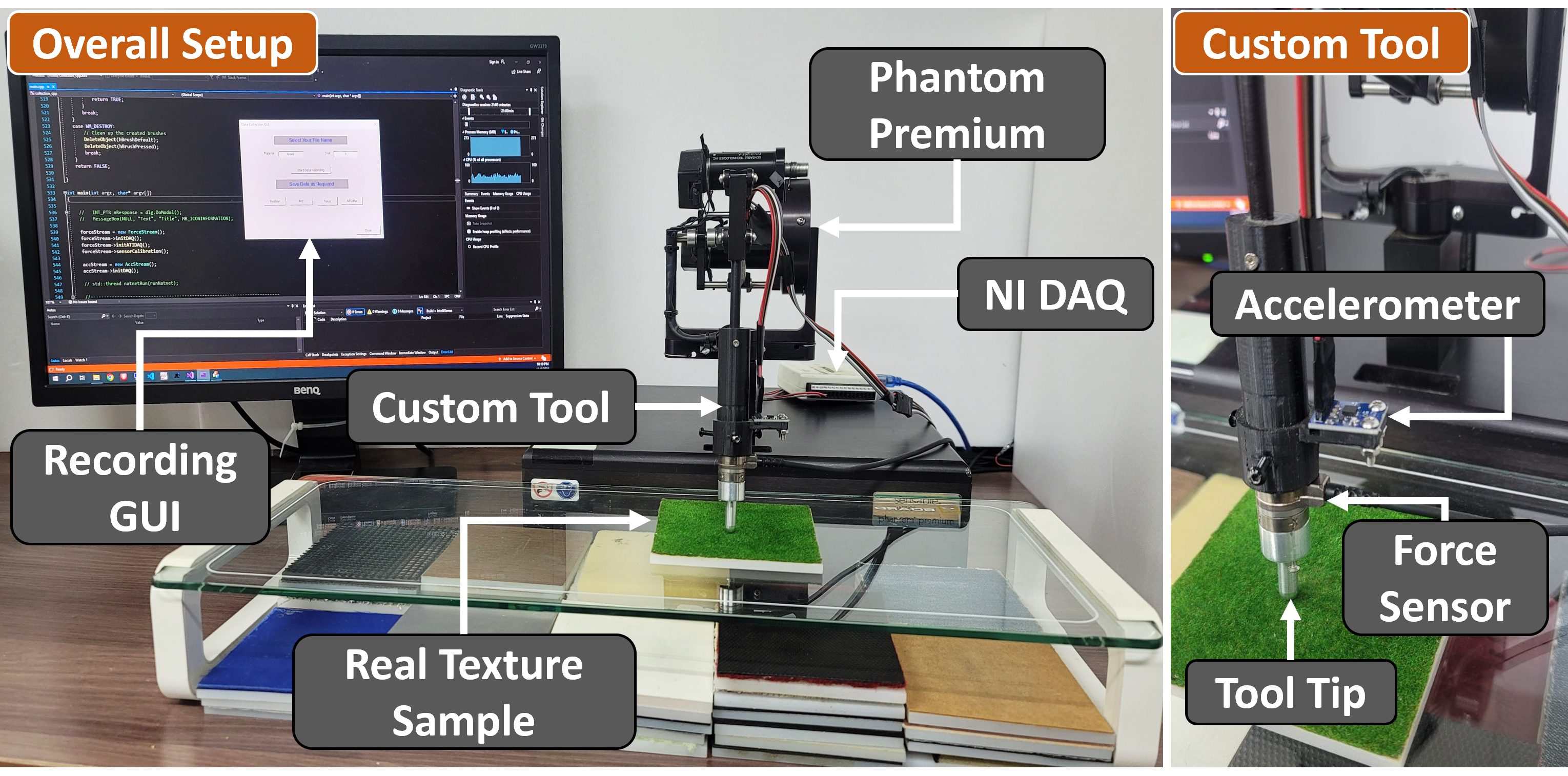}
        \caption{Data recording setup. The setup records vibrations produced when the user rubs the surface, along with the applied speed and force.}
        \label{fig:Tactile_Data_Recording_Setup}
\end{figure}

\textbf{Apparatus:} The tactile data acquisition setup is illustrated in Figure~\ref{fig:Tactile_Data_Recording_Setup}. It consists of a rigid tool equipped with a detachable 2.0 mm hemispherical stainless steel tip. The tool body is custom-designed and fabricated using ABS plastic.
A 3-axis accelerometer (ADXL335, Analog Devices) is mounted on the tool to record vibrations during surface exploration, while a force sensor (Nano17, ATI Industrial Automation) measures forces along three axes. The tool is mounted on a Phantom Premium haptic device, enabling precise tracking of position and orientation for accurate speed at 1 kHz and normal force estimation.
The accelerometer connects to a PC via a data acquisition card (USB-6351, National Instruments), recording at 3 kHz. The force sensor uses a dedicated DAQ system to sample forces at 8 kHz. This hardware configuration is consistent with setups commonly used in haptic research for texture data collection and offers high-resolution measurements suitable for tactile signal analysis~\cite{abdulali2016data}.

\begin{figure} [t]
\centering
        \includegraphics[width=1\columnwidth] {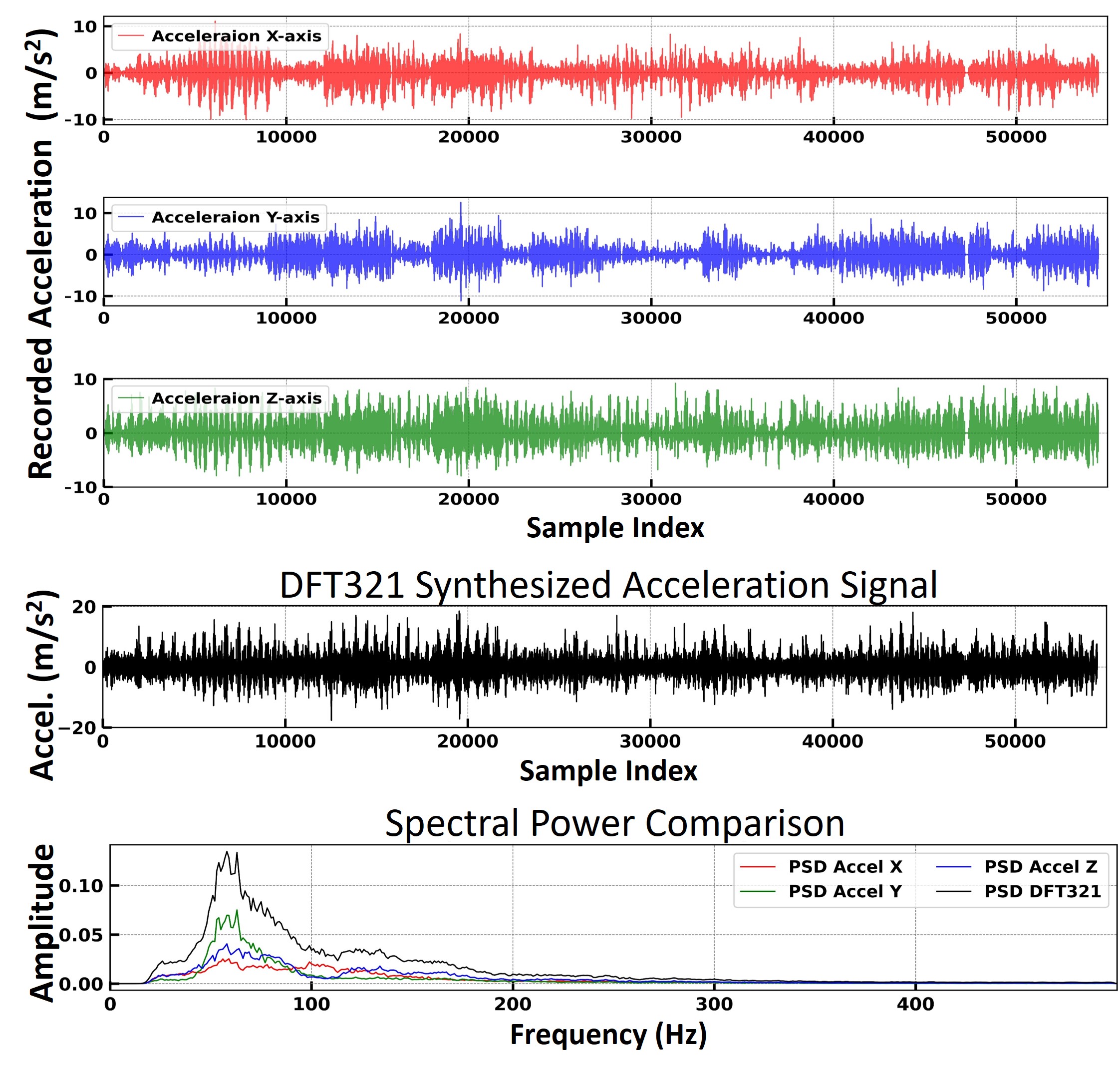}
        \caption{Acceleration signals for artificial grass recorded along three axes (first three plots) and combined into a single-axis using the DFT321 algorithm (fourth plot), retaining temporal characteristics and spectral power (fifth plot).}
        \label{fig:Acceleration_signals}
\end{figure}

\vspace{1em}
\noindent \textbf{Data Collection and Pre-processing:} 
Interaction data was collected for all 50 textures detailed in Sec. \ref{subsec:texture_dataset}. Each texture was recorded for 60 seconds using freehand motion to capture natural surface interactions.
All recorded data was resampled at 1000 Hz for uniformity. The initial and final 2.5 seconds were cropped to reduce artifacts and eliminate stationary effects. Interaction signals, including scanning speed and normal force, were low-pass filtered at 25 Hz to suppress high-frequency noise, while acceleration signals were band-pass filtered between 20 Hz and 500 Hz to isolate relevant vibrations and remove gravitational components \cite{abdulali2016data,culbertson2014modeling}.
The 3-axis acceleration signals were projected onto a single axis using the DFT321 algorithm, preserving temporal and spectral characteristics \cite{landin2010dimensional}. Scanning speed was derived by combining velocities along all three axes, and normal force was computed by projecting 3-axis force vectors onto the surface normal.
Figure \ref{fig:Acceleration_signals} shows the final processed data for artificial grass texture (T50).

\begin{figure} [t]
\centering
        \includegraphics[width=0.98\columnwidth] {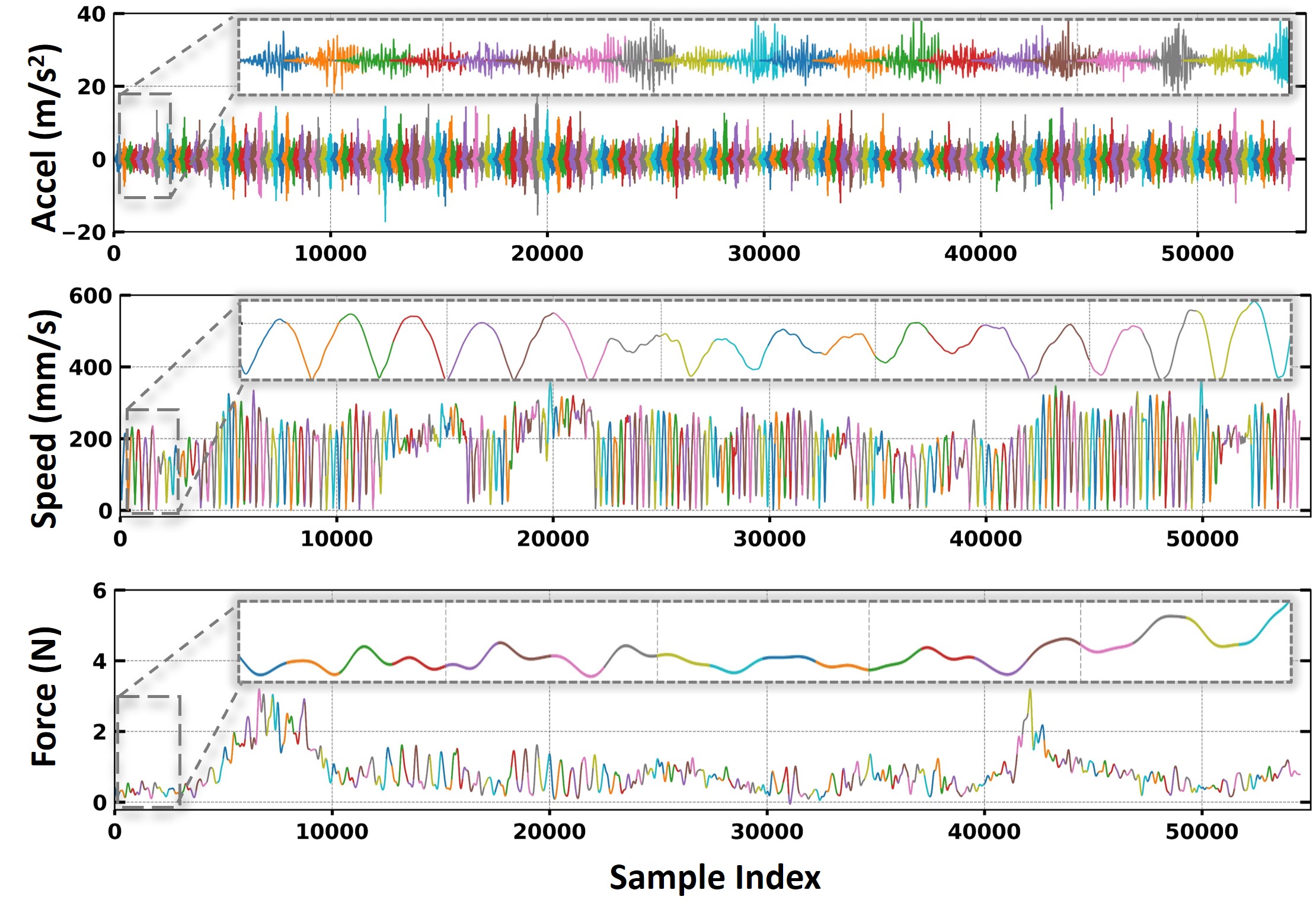}
       \caption{The processed acceleration signal along with interaction signals, including speed and force. Each graph also shows the segments created during pre-processing.}

        \label{fig:Segmented_signasl}
\end{figure}

\vspace{1em}
\noindent \textbf{Mel Frequency Cepstral Coefficients (MFCC):}
Physical acceleration signals collected from textured surfaces contain valuable haptic information alongside redundant data. To extract meaningful haptic features from these signals, we apply Mel Frequency Cepstral Coefficients (MFCC), a technique widely used in audio and signal processing \cite{kim2006mpeg}. MFCC effectively captures essential vibrational patterns from texture data, making it well-suited for haptic analysis. It has been successfully applied to surface classification and texture modeling \cite{strese2015surface, hassan2019authoring}.

To compute MFCCs, the raw acceleration signals were divided into 0.5-second segments (500 samples at 1000 Hz). Each segment was windowed using a Hann window and further divided into 25 ms frames with 50\% overlap, resulting in 40 frames per segment. For each frame, 13 MFCC coefficients were extracted, producing a matrix of size 40 × 13. Flattening this matrix yielded 520 MFCC features per segment.

On the other hand, speed and force signals are low-frequency and exhibit minimal variation over short intervals; their minimum, maximum, and average values were computed for each segment, contributing 6 additional features. The final feature vector, comprising 526 features per segment, was computed using Python’s SciPy and librosa libraries. Each segment's features serve as input to the tactile network (HT-Net). An illustration of the segmented acceleration, speed, and force signals for Artificial Grass surface (T50) is shown in Figure \ref{fig:Segmented_signasl}.

\subsection{Image Dataset }
\label{subsec:Image_dataset_collection}

The proposed multi-modal strategy incorporates texture images to extract visual features for predicting haptic ratings. Traditionally, haptic applications have used raw images with classical texture descriptors like Gray-Level Co-occurrence Matrix (GLCM), Gabor filters, and Local Binary Patterns (LBP) \cite{strese2016multimodal}. Recently, pre-trained deep learning models have become popular for visual feature extraction in texture analysis and tactile perception \cite{lin2024object}.
Despite their effectiveness, deep learning models can miss surface details when target textures differ from those in the training datasets. To mitigate this, we adopted a hybrid approach combining classical and deep learning-based feature extraction. GLCM, a robust texture descriptor, was used alongside features from a pre-trained deep learning model.

\vspace{1em}
\noindent \textbf{Image Capturing Setup:}
Learning haptic properties from images requires capturing fine surface details and granularity. To achieve this, high-resolution images are essential. We developed a setup using a dp2 Quattro SIGMA camera mounted on a tripod, maintaining a fixed distance of 30 cm between the camera lens and the surface.
For each of the 50 textures, 10 images were captured under varying lighting and angular conditions to enhance feature diversity and generalization. To minimize boundary blur, all images were cropped and resized to 1568 X 1568 from the center. 

\vspace{1em}
\noindent \textbf{DL-Based Features:}
ResNet \cite{he2016deep}, known for its deep architecture and residual connections, has demonstrated significant performance in image classification and feature extraction tasks. Its ability to capture fine-grained details makes it well-suited for applications requiring dense visual representations, including haptic texture analysis \cite{lin2024object}.

In this study, we employed ResNet-50\cite{he2016deep}, pre-trained on ImageNet, to extract feature vectors, a method validated in prior haptic research \cite{ lin2024object, hassan2023establishing}. To maintain the resolution of texture images and avoid loss of detail, each image was divided into 49 overlapping 224×224 patches, matching ResNet's input size.
For each patch, feature vectors of size 1×2048 were extracted from the average pooling layer. These vectors were averaged across all patches to generate the final feature representation for each image.
Notably, since texture perception is less dependent on color, all images were converted to grayscale. To ensure compatibility with ResNet’s three-channel input, grayscale images were replicated across three channels, allowing seamless use of the pre-trained model without modifying its architecture.

\vspace{1em}
\noindent \textbf{Classical Texture Descriptors:}
For classical texture analysis, we employed the Gray-Level Co-occurrence Matrix (GLCM) \cite{haralick1973textural}, a widely used method for capturing spatial relationships between pixel intensities. GLCM has also been extensively applied in texture analysis and has shown significant success in haptic studies for characterizing surface properties \cite{hassan2023establishing}. In this study, the GLCM was computed on surface texture images quantized to 16 gray levels, resulting in a 16×16 matrix. This matrix was then flattened to generate a feature vector of size 1×256. 

In the final stage, the extracted GLCM features were combined with deep learning (DL)-based features obtained from ResNet-50. The 2048-dimensional feature vectors were extracted from the average pooling layer of ResNet-50 and concatenated with the GLCM features, resulting in a comprehensive image-based feature vector of size 1 × 2304. This combined feature vector served as the input to the Haptic Vision Network (HV-Net) for further processing.
It is noted that, to improve generalization and better align the visual input space with the temporal tactile segments, visual samples were dynamically augmented during training using TensorFlow’s data pipeline. The augmentation process included random rotations, horizontal and vertical flips, and Gaussian noise, thereby enriching visual diversity across input conditions.

\section{Evaluation Experiments}
\label{sec:evaluation_experiments}

The primary objective of this experiment is to evaluate the effectiveness of the proposed approach in estimating perceptual attributes of textured surfaces. The following sections outline the error metrics, the leave-one-out cross-validation (LOOCV) technique for unseen data, and the results obtained. The framework is compared with existing methods, followed by an analysis of different feature sets.

\subsection{Error Metrics }
\label{sec:error_metric}

To evaluate the performance of the proposed framework, we employed Mean Absolute Error (MAE) and Root Mean Square Error (RMSE) as the primary error metrics. MAE quantifies the average magnitude of errors, while RMSE penalizes larger deviations, providing a comprehensive measure of prediction accuracy. These metrics were used to assess the model’s effectiveness in estimating individual haptic attribute estimation accuracy by comparing predicted values with user-provided ratings. The attributes evaluated are the same as those discussed in Sec. \ref{sec:haptic_perceptual_space}, including Rough-smooth (R-S), Flat-bumpy (F-B), Sticky-slippery (S-S), and Hard-soft (H-S).
These metrics are widely used in related studies \cite{awan2023predicting,hassan2023establishing} and are defined as follows:

\begin{equation}
\text{MAE} = \frac{1}{n} \sum_{i=1}^{n} |y_{i} - \tilde{y}_{i}|,
\label{eq:mae}
\end{equation}

\begin{equation}
\text{RMSE} = \sqrt{\frac{1}{n} \sum_{i=1}^{n} (y_{i} - \tilde{y}_{i})^2},
\end{equation}

where \( y_{i} \) represents the actual rating provided by the user for the \( i^{\text{th}} \) texture sample, \( \tilde{y}_{i} \) denotes the estimated attribute rating, and \( n \) is the total number of observations or texture samples. 
It is important to note that the actual and predicted values are scaled to the range of 0 to 100 before computing the error. 
For example, an MAE of 10 represents an average deviation of 10 on a 100-point scale, reflecting the difference between predicted and actual user ratings.

\begin{figure*} [t]
\vspace{-0.5cm}
\centering
        \includegraphics[width=1.0\textwidth, height = 8.5cm] {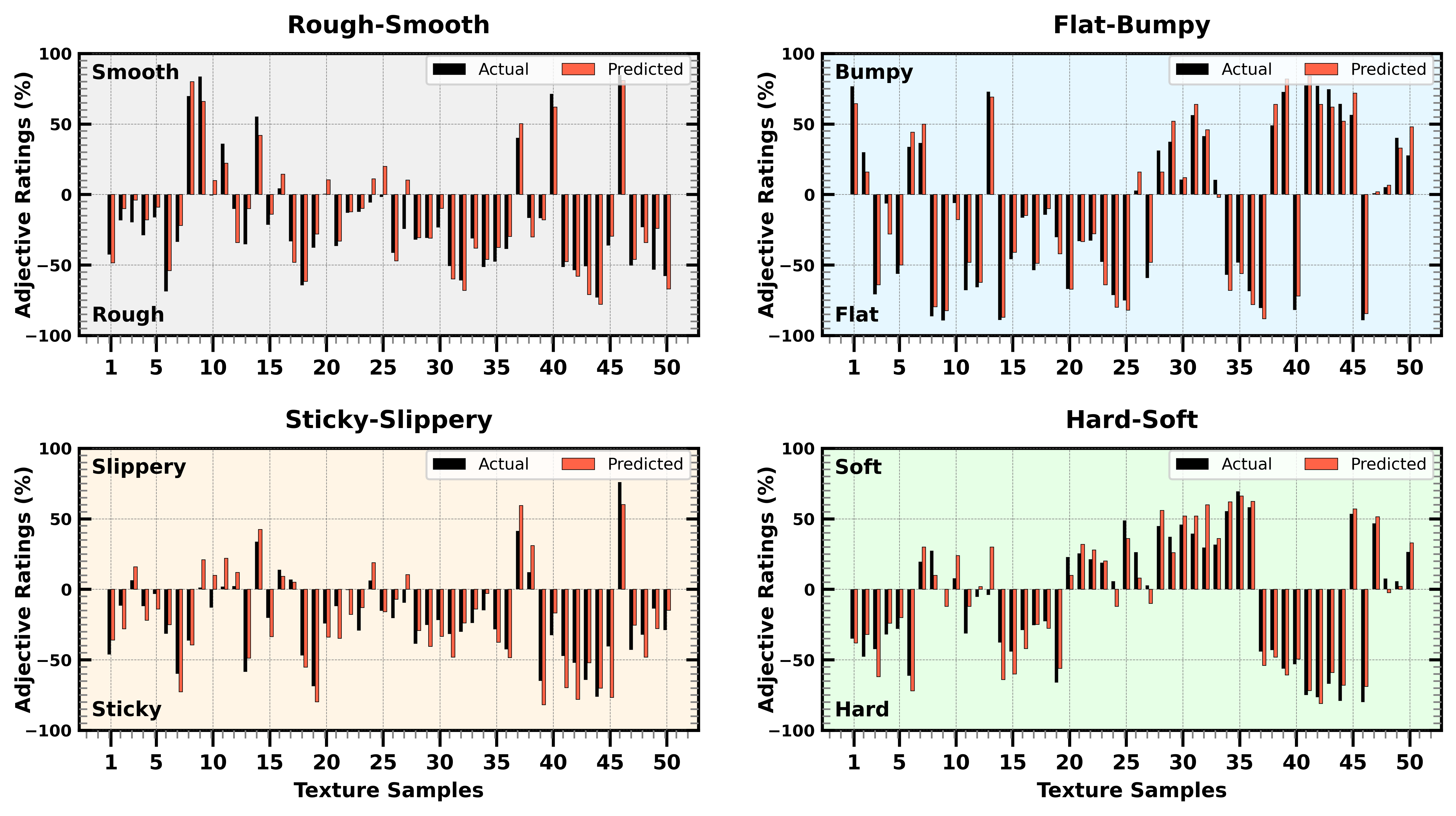}
        \caption{Comparison of actual and predicted attributes for 50 textures using the Leave-One-Out Cross-Validation (LOOCV) technique.}

        \label{fig:loocv_results}
\end{figure*}

\subsection{Leave-One-Out Cross Validation (LOOCV)}  
\label{subsec:loocv}  

Fitting high-dimensional data requires robust validation techniques to identify the most optimized models and ensure reliable performance across different methods. One of the most widely used validation approaches is cross-validation, which evaluates a model’s ability to generalize by repeatedly training and testing it on different subsets of data \cite{patil2021uniform}.  
Among the various types of cross-validation, k-fold cross-validation is the most common. In this approach, the dataset is divided into \( k \) equally sized subsets (typically \( k = 5 \) or \( k = 10 \)). During each iteration, one subset is held out for validation, while the remaining \( k-1 \) subsets are used for training. This process repeats \( k \) times, with each subset serving as the validation set once. The model’s final performance is averaged across all iterations. However, choosing a small \( k \) (such as 5 or 10) can sometimes lead to high bias and underfitting, particularly when dealing with small datasets.  
To address this, leave-one-out cross-validation (LOOCV), a special case of k-fold cross-validation where \( k = n \) (the total number of observations), is often employed. In LOOCV, the data is split into \( n \) subsets, where each iteration uses a single data point for validation while the remaining \( n-1 \) points are used for training . This process repeats for every observation, ensuring that each data point is tested exactly once, which significantly reduces bias and leverages the entire dataset for model training \cite{patil2021uniform,lumumba2024comparative,stone1974cross}.  

In this work, we utilized LOOCV to evaluate the performance of the proposed framework. Our dataset consists of 50 textures (\( n=50 \)). For each iteration, 49 textures (\( n-1 \)) were used for training, while the remaining texture was reserved for validation. This resulted in 50 training cycles, providing a comprehensive assessment of the model's generalizability. Despite being computationally demanding, LOOCV is particularly valuable for small datasets, as it maximizes the use of available data and yields reliable performance estimates on unseen samples. This makes it an ideal choice for our evaluation process.

\subsection{Model Performance}
\label{subsec:model_performance}

Figure \ref{fig:loocv_results} presents the comparison between actual and predicted values from the proposed visuo-tactile Net for each texture. The results are plotted within a range of -100 to 100 for each surface. As shown, the model's predictions align closely with user ratings for most textures.  

To further assess accuracy, we calculated MAE across all attributes. The lowest error of 4.48 was recorded for the F-B attribute, followed by H-S and R-S with 5.21 and 5.23, respectively. The highest error of 6.67 was observed for the S-S attribute, as shown in Table \ref{tab:mae_compariosn_with_others}. All MAE values are scaled from 0 to 100, as described in Sec. \ref{sec:error_metric}, ensuring consistent interpretation across different attributes.
Additionally, class-based errors are visualized in Figure \ref{fig:heatmap}. The results show that paper and jeans textures exhibited the highest errors, whereas other texture classes achieved stronger predictive performance. Further details are provided in Sec. \ref{sec:discussion}.

\begin{table} [b]
\caption{Mean Absolute Error (MAE) values for the proposed system and five algorithms across four attribute pairs.}
\label{tab:mae_compariosn_with_others} 
\centering
\renewcommand{\arraystretch}{1.1} 

\begin{tabular}{l c c c c}
\hline
\hline
\noalign{\vskip 1.1mm} 

\textbf{Methods} & \textbf{R-S} & \textbf{F-B} & \textbf{S-S} & \textbf{H-S} \\
                
\noalign{\vskip 1.1mm} 

\hline
\hline
\noalign{\vskip 1.1mm} 

Artificial Neural Network  & 21.13 & 26.12 & 22.85 & 25.44\\
\noalign{\vskip 1.1mm}

Vision 1D-CNN \cite{taye2020application}  & 18.55 & 19.63 & 17.89 & 17.24\\
\noalign{\vskip 1.1mm}

Haptic CNN\cite{hassan2023establishing}          &13.17  & 11.32 & 12.01 &  8.38 \\
\noalign{\vskip 1.1mm} 

Tactile CNN-LSTM \cite{awan2023predicting}  & 10.58 & 8.98 & 13.76 & 11.92 \\
\noalign{\vskip 1.1mm} 

Tactile SVM\cite{shao2023haptic}                 & 9.40 & 14.89 & 15.35 & 10.54 \\
\noalign{\vskip 1.1mm}

\hline
\noalign{\vskip 1.1mm} 

\textbf{Proposed Method}            & \textbf{5.23} & \textbf{4.48} & \textbf{6.67} & \textbf{5.21} \\
\noalign{\vskip 1.1mm} 
\hline

\end{tabular}
\end{table}

\subsection{Comparison with Baseline Models}
\label{subsec:comparison_baseline}

The performance of the proposed framework was evaluated against other similar strategies for estimating haptic texture attribute ratings using either visual and/or tactile data. These include Vision 1D-CNN~\cite{taye2020application}, Haptic CNN~\cite{hassan2023establishing}, Tactile CNN-LSTM~\cite{awan2023predicting}, Tactile SVM~\cite{shao2023haptic}, and a multimodal artificial neural network (ANN) baseline. All models were implemented using TensorFlow 2.7, and their core architectures were reproduced based on the original publications. For consistency, the final regression layer of each model was modified to produce four continuous outputs corresponding to the four haptic attributes.
The ANN baseline provides a simplified multimodal fusion benchmark without explicit modeling of spatial or temporal structure. It uses the same extracted features as the proposed framework: flattened ResNet and GLCM features for vision, and MFCC with statistical descriptors for tactile input. These are passed through two parallel fully connected branches (layer sizes: 128, 256, 256, 128), followed by feature fusion and two regression layers of 64 units. A final dense layer outputs four predicted values. This setup allows examination of the benefits introduced by modality-specific and structured processing.
 
The results, shown in Table \ref{tab:mae_compariosn_with_others} for MAE and Table \ref{tab:rmse_compariosn_with_others} for RMSE, demonstrate that the proposed method consistently outperforms baseline models across all attribute pairs. The proposed model achieved the lowest errors in both MAE and RMSE, reflecting its superior accuracy and generalizability.
For MAE, the proposed method recorded values of 5.23 for R-S, 4.48 for F-B, 6.67 for S-S, and 5.21 for H-S. In contrast, the ANN exhibited significantly higher errors, with 21.13 for R-S and 25.44 for H-S. Similar trends were observed in RMSE, where the proposed model achieved the lowest errors at 6.81 (R-S), 5.67 (F-B), 7.52 (S-S), and 6.13 (H-S). The ANN, by comparison, yielded RMSE values of 24.41 (R-S) and 29.12 (H-S).  

Among the baseline models, Tactile SVM \cite{shao2023haptic} and CNN-LSTM \cite{awan2023predicting} outperformed ANN but remained less accurate than the proposed method. For the F-B attribute, the proposed model achieved a lower MAE of 4.48 compared to 18.55 from Vision 1D-CNN \cite{taye2020application}. A similar trend appeared in RMSE, where \cite{taye2020application} produced an error of 24.85, while the proposed method achieved a significantly lower RMSE of 5.67. Interestingly, vision-based models \cite{taye2020application} and \cite{hassan2023establishing} consistently produced higher errors compared to tactile-based approaches, highlighting the advantage of tactile data for haptic attribute estimation and the strength of the proposed  visuo-tactile multi-model based technique in outperforming vision-based or tactile-based approaches.

\subsection{Individual Feature Error}
\label{subsec:error_analysis}

In this section, we evaluate the performance of individual feature extraction techniques for both visual and tactile data, as well as the benefits of combining them. The goal is to identify which features contribute most to reducing errors in haptic attribute estimation. We assess features from ResNet-50 and GLCM for vision \cite{lu2022preference,hassan2023establishing}, and 1D Discrete Wavelet Transform (1D-DWT), Discrete Fourier Transform (DFT), and MFCC for tactile data. These features were selected based on their effectiveness in haptic contexts \cite{slepyan2024wavelet, heravi2024development, strese2016multimodal}.

Table~\ref{tab:feature_comparison} presents the performance of individual and combined features across both visual and tactile modalities. For vision-based inputs, concatenating ResNet and GLCM features led to improved accuracy across all attributes. The combined visual features achieved an RMSE of 10.11 for R-S, outperforming ResNet (18.29) and GLCM (19.11) individually. Similar improvements were observed for F-B and S-S. On the tactile side, MFCC consistently outperformed 1D-DWT and DFT, achieving an RMSE of 9.89 for R-S compared to 31.3 and 34.61, respectively. Notably, due to the poor performance of DWT and DFT, their combination with MFCC was not pursued, as initial trials led to unstable results and degraded performance. Combining visual and tactile features further reduced errors, resulting in the lowest RMSE across most attributes. The proposed model, integrating ResNet, GLCM, and MFCC, achieved RMSE values of 6.81 for R-S and 5.67 for F-B. Tactile data alone (MFCC, 11.35) also outperformed vision-only features for F-B, emphasizing the importance of tactile input for certain perceptual dimensions. Overall, the findings demonstrate the effectiveness of multi-feature, multimodal fusion in improving haptic attribute prediction.

\begin{table} [t]
\caption{Root Mean Square Error (RMSE) values for the proposed system and five algorithms across four attribute pairs.}
\label{tab:rmse_compariosn_with_others} 
\centering
\renewcommand{\arraystretch}{1.1} 

\begin{tabular}{l c c c c}
\hline
\hline
\noalign{\vskip 1.1mm} 

\textbf{Methods} & \textbf{R-S} & \textbf{F-B} & \textbf{S-S} & \textbf{H-S} \\
                
\noalign{\vskip 1.1mm} 

\hline
\hline
\noalign{\vskip 1.1mm} 

Artificial Neural Network  & 24.41 & 31.62 & 25.73 & 32.19\\
\noalign{\vskip 1.1mm}

Vision 1D-CNN \cite{taye2020application}  & 22.35 & 24.88 & 19.61 & 20.59\\
\noalign{\vskip 1.1mm} 

Haptic CNN\cite{hassan2023establishing}                 & 18.21 & 12.15 & 14.19 &  12.65
\\
\noalign{\vskip 1.1mm}

Tactile CNN-LSTM \cite{awan2023predicting}  & 13.45 & 10.65 & 15.20 & 13.78 \\
\noalign{\vskip 1.1mm} 

Tactile SVM\cite{shao2023haptic}                 & 11.26 & 16.37 & 20.81 & 11.93 \\
\noalign{\vskip 1.1mm}

\hline
\noalign{\vskip 1.1mm} 

\textbf{Proposed Method}        & \textbf{6.81} & \textbf{5.67} & \textbf{7.52} & \textbf{6.13} \\
\noalign{\vskip 1.1mm} 
\hline

\end{tabular}
\end{table}

\begin{table} [t]
\caption{RMSE of individual features compared to concatenated features.}
\label{tab:feature_comparison} 
\centering
\renewcommand{\arraystretch}{1.0} 

\begin{tabular}{l c c c c c}
\hline
\hline
\noalign{\vskip 1.0mm} 

\textbf{Feature Type} & \textbf {Feature} & \textbf{R-S} & \textbf{F-B} & \textbf{S-S} & \textbf{H-S} \\
                
\noalign{\vskip 1.0mm} 

\hline
\hline
\noalign{\vskip 1.0mm} 

Vision          & ResNet & 18.29 & 16.52 & 15.36 & 13.50\\
\noalign{\vskip 1.0mm} 

                & GLCM& 19.11 & 12.53 & 10.14 & 14.96\\
\noalign{\vskip 1.0mm} 

                & Concatenated& 13.26 & 10.11 & 12.52 & 8.6 \\
\noalign{\vskip 1.0mm} 

\hline
\noalign{\vskip 1.0mm} 

Tactile        & 1D-DWT & 31.3 & 46.8 & 42.5 & 39.3\\
\noalign{\vskip 1.0mm} 

                & DFT & 34.61 & 29.85 & 26.51 & 28.41\\
\noalign{\vskip 1.0mm} 

                & MFCC & 9.89 & 11.35 & 10.71 & 7.98\\
\noalign{\vskip 1.0mm} 

\hline
\noalign{\vskip 1.0mm} 

\textbf{Proposed}      & \textbf{ResNet+GLCM} & \textbf{6.81} & \textbf{5.67} & \textbf{7.52} & \textbf{6.13} \\
\textbf{Method} & \textbf{MFCC} &&  &  & \\

\hline

\end{tabular}
\end{table}

\section{Discussion} 
\label{sec:discussion}

Building on the findings presented in Figure~\ref{fig:loocv_results} and Table~\ref{tab:mae_compariosn_with_others}, this section examines attribute-wise prediction trends, modality-specific behavior, and class-level error patterns to better understand the strengths and limitations of the proposed framework. Among the four attribute pairs, S-S exhibited the highest error, while F-B achieved the lowest, as reflected by the average MAE and RMSE. R-S and H-S showed moderate errors, performing better than S-S but not as accurately as F-B. 

Notably, considering the effect of visual and tactile features, we found that each modality has its strengths, and their combination yields superior results. As shown in Table \ref{tab:feature_comparison}, the vision-based approach performed better in capturing the flat-bumpy (F-B) attribute compared to the tactile-based approach. This may be due to the visual features' ability to clearly detect surface patterns, while tactile signals, particularly acceleration data, may introduce noise during deep strokes, resulting in undesired bounciness.

\begin{figure} [t]
\centering
        \includegraphics[width=1\columnwidth] {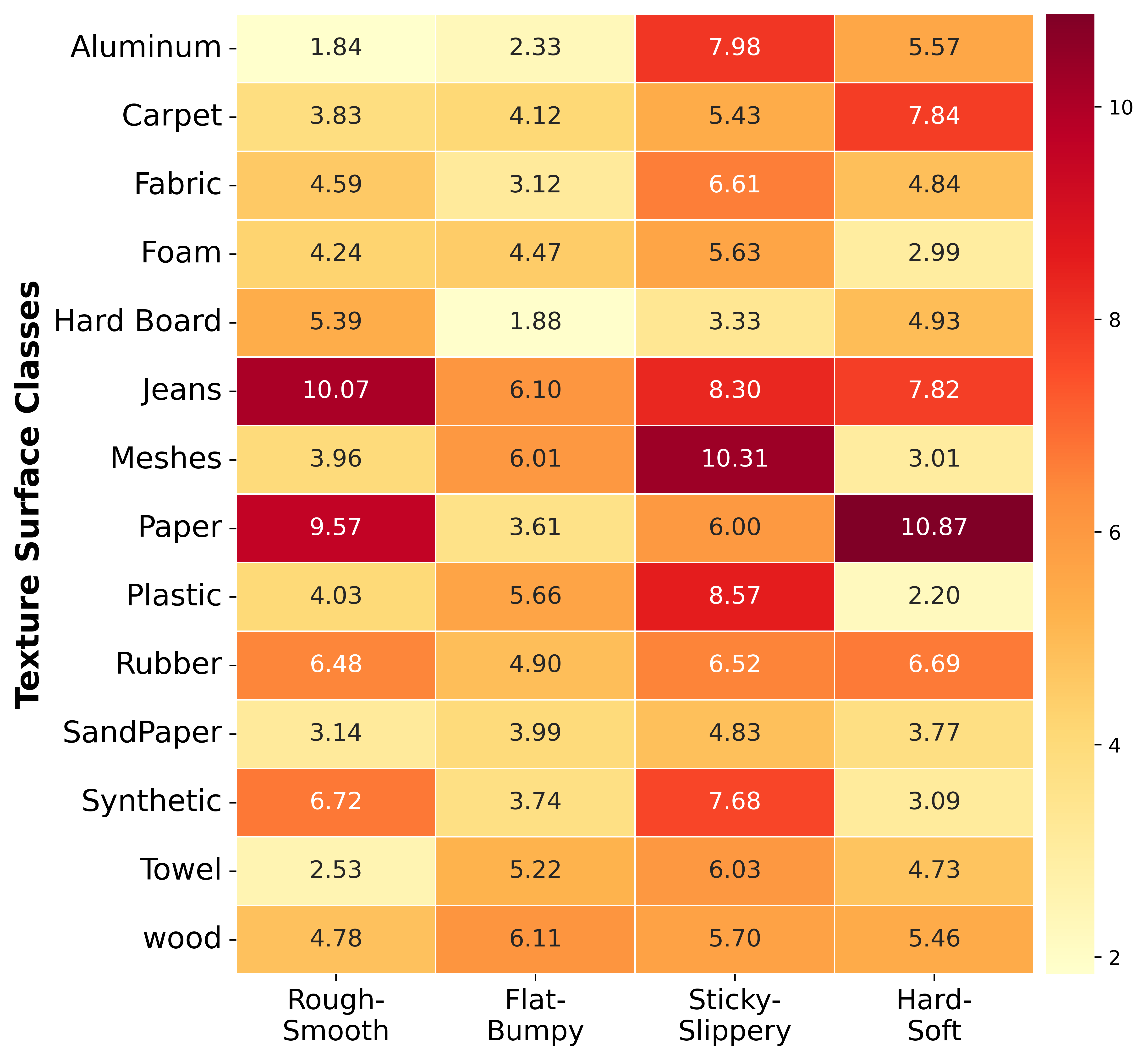}
        \caption{Heatmap of MAE for four haptic attribute pairs across various texture classes. Darker shades indicate higher errors, while lighter shades show lower errors, reflecting prediction performance across different texture classes using the visuo-tactile Net.}
        \label{fig:heatmap}
\end{figure}

Figure \ref{fig:heatmap} highlights performance variations across texture classes, with paper and jeans categories exhibiting the highest errors across most attribute pairs. For paper textures, the highest MAE was recorded for H-S at 10.87 and R-S at 9.57, likely due to the diverse range of samples, including both plain and heavily textured surfaces. Since the model maps the physical signal space to a perceptual space derived from human ratings, it is plausible that participants may have overlooked finer details. Perceptual biases driven by preconceived judgments, as noted in \cite{hassan2016evaluating}, could have influenced ratings, where participants assess haptic qualities based on prior experiences rather than the actual textures presented during the experiment.
A similar pattern emerged in the jeans category, particularly for T27 (smooth jeans), where surface texture variations likely contributed to increased errors, reflecting challenges akin to those encountered with paper textures. Additionally, the meshes class showed elevated errors in the S-S attribute (MAE 10.31), which may be attributed to noise artifacts accumulating during tactile data recording. The rigid and structured nature of hard plastic and metal meshes could have introduced inconsistencies, resulting in higher prediction errors.
Despite these discrepancies, the errors remain within acceptable bounds, aligning with the Just Noticeable Difference (JND) threshold for perceptual similarity, often estimated at around 10 out of 100 \cite{hassan2024quantifying}. Most class-wise and overall average MAE values fall below this threshold, reinforcing the effectiveness of the proposed framework.

The generalizability of the model is further demonstrated by its performance on unique textures such as aluminum (T46), which exhibits distinct surface properties. Despite its uniqueness, aluminum performed well, with the highest error recorded in the sticky-slippery (S-S) attribute at 7.98. This elevated error may be attributed to rubbing marks left by the interaction tool, a known phenomenon in tactile studies. Since aluminum is the sole sample in its category, further investigation is necessary to better understand this behavior. We believe that incorporating additional textures with similar properties will enhance overall performance. However, it can be argued that the study has yet to encounter a sufficiently diverse range of textures.
Expanding the dataset with additional textures is likely to improve texture attribute prediction quality. Although LOOCV can introduce biases in certain cases, it remains an effective method for comprehensive evaluation. The results clearly indicate that the proposed autoencoder-based framework, combined with CNN and feature-based inputs, captures nuanced surface properties and represents an improvement over existing single-modality approaches.

\section{Conclusion}
\label{sec:conclusion}

This study introduces a deep learning visuo-tactile framework for predicting haptic texture attributes. It maps a physical signal space, constructed from visual and tactile features, to a perceptual space defined by user ratings. The four-dimensional perceptual space includes the bipolar pairs: rough–smooth, flat–bumpy, hard–soft, and sticky–slippery. The architecture combines a CNN-based autoencoder for visual processing with a ConvLSTM network for modeling tactile signal dynamics. Visual inputs are encoded using features from ResNet and GLCM, while tactile signals are represented using MFCCs derived from high-frequency acceleration data. The framework demonstrates improved prediction accuracy over existing methods by integrating visual and tactile data in a unified manner. These results confirm the framework's reliability and scalability in estimating haptic attributes, with potential utility in material recognition when user ratings are unavailable or difficult to collect, assisting researchers in assessing perceptual responses, and in robotic perception systems where accurate surface interpretation is essential.

\bibliographystyle{IEEEtran}  
\bibliography{references}     

\begin{IEEEbiography}[{\includegraphics[width=1in,height=1.25in,clip,keepaspectratio]{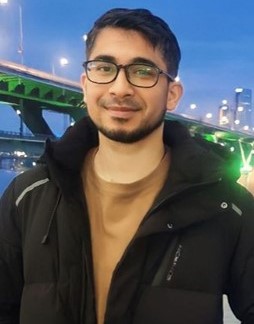}}]
{Mudassir Ibrahim Awan} received his B.E. in Electronics Engineering from the Karachi Institute of Economics and Technology (KIET), Karachi, Pakistan, in 2016. 
In 2018, he joined the Haptics and Virtual Reality Laboratory at Kyung Hee University, South Korea, where he is currently pursuing an integrated MS-PhD program in the Department of Computer Science and Engineering. His research interests include haptic modeling and rendering, psychophysics, drone haptics,  and car door perception.
\end{IEEEbiography}

\begin{IEEEbiography}
[{\includegraphics[width=1in,height=1.25in, clip,keepaspectratio]{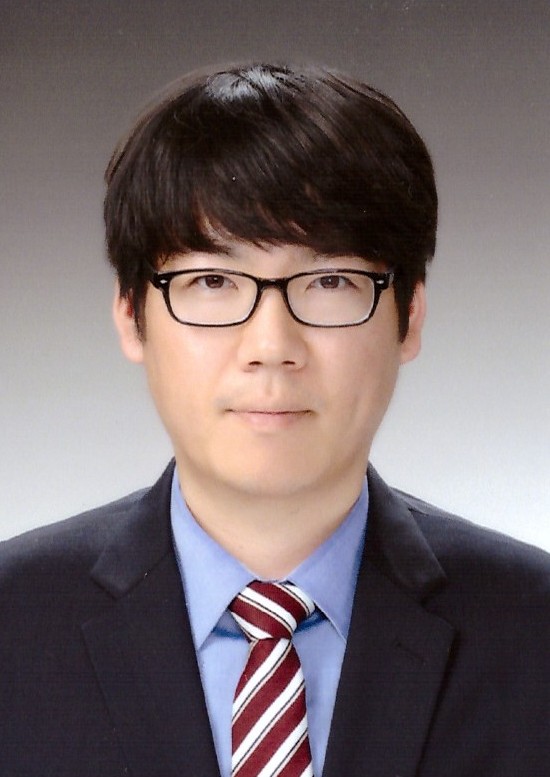}}]{Seokhee Jeon} He received his B.S. and Ph.D. degrees in Computer Science and Engineering from Pohang University of Science and Technology (POSTECH) in 2003 and 2010, respectively. He then worked as a Postdoctoral Research Associate at the Computer Vision Laboratory, ETH Zurich. In 2012, he joined the Department of Computer Engineering at Kyung Hee University as an Assistant Professor and became a Full Professor in 2024. He is also a co-founder and faculty member in the Department of Metaverse.
His research interests include data-driven haptic modeling and rendering, hyper-realistic multimodal feedback in virtual, augmented, and remote environments, and the development of modular wearable haptic interfaces with enhanced applicability.
\end{IEEEbiography}

\EOD

\end{document}